\documentclass[twocolumn, nolinenumbers]{aastex631}

\usepackage{appendix}
\usepackage{graphicx}
\usepackage{amsmath}

\begin{document}

\title{Interstellar carbonaceous dust erosion induced by X-ray irradiation of water ice in star-forming regions}
\shorttitle{X-ray irradiation of H$_2$O on $^{13}$C dust}

\author[0000-0001-6877-5046]{K.-J. Chuang}
\affiliation{Laboratory for Astrophysics, Leiden Observatory, Leiden University, P.O. Box 9513, NL-2300 RA Leiden, the Netherlands.}
\email{chuang@strw.leidenuniv.nl}

\author[0000-0001-7803-0013]{C. Jäger}
\affiliation{Laboratory Astrophysics Group of the Max Planck Institute for Astronomy at the Friedrich Schiller University Jena, Institute of Solid State Physics, Helmholtzweg 3, D-07743 Jena, Germany}

\author[0000-0001-7755-7884]{N.-E. Sie}\altaffiliation{Current address: Institute of Low Temperature Science, Hokkaido University, Sapporo, Hokkaido 060-0819, Japan}
\affiliation{Department of Physics, National Central University, Zhongli Dist., Taoyuan City 320317, Taiwan}

\author[0000-0003-2741-2833]{C.-H. Huang}\altaffiliation{Current address: National Synchrotron Radiation Research Center, 101 Hsin-Ann Road, Hsinchu Science Park, Hsinchu 30076, Taiwan}
\affiliation{Department of Physics, National Central University, Zhongli Dist., Taoyuan City 320317, Taiwan}


\author{C.-Y. Lee} 
\affiliation{Department of Physics, National Central University, Zhongli Dist., Taoyuan City 320317, Taiwan}

\author{Y.-Y. Hsu} 
\affiliation{Department of Physics, National Central University, Zhongli Dist., Taoyuan City 320317, Taiwan}

\author[0000-0002-1493-300X]{Th. Henning}
\affiliation{Max Planck Institute for Astronomy, Königstuhl 17, D-69117 Heidelberg, Germany}

\author[0000-0003-4497-3747]{Y.-J. Chen}
\affiliation{Department of Physics, National Central University, Zhongli Dist., Taoyuan City 320317, Taiwan}
\email{asperchen@phy.ncu.edu.tw}



\begin{abstract}

The chemical inventory of protoplanetary midplanes is the basis for forming planetesimals. Among them, solid-state reactions based on CO/CO$_2$ toward molecular complexity on interstellar dust grains have been studied in theoretical and laboratory work. 
The physicochemical interactions between ice, constituted mainly of H$_2$O, and dust surfaces are limited to a few experimental studies focusing on vacuum ultraviolet and cosmic-ray processing.
In this work, the erosion of C dust grains induced by X-ray irradiation of H$_2$O ice was systematically investigated for the first time. The work aims to provide a better understanding of the reaction mechanism using selectively isotope-labeled oxygen/carbon species in kinetic analysis. 
Ultrahigh vacuum experiments were performed to study the interstellar ice analog on sub-$\mu$m thick C dust at $\sim$13~K. H$_2$O or O$_2$ ice was deposited on the pre-synthesized amorphous C dust and exposed to soft X-ray photons (250--1250~eV). Fourier-transform infrared spectroscopy was used to monitor in situ the newly formed species as a function of the incident photon fluence. Field emission scanning electron microscopy was used to monitor the morphological changes of (non-)eroded carbon samples.
The X-ray processing of the ice/dust interface leads to the formation of CO$_2$, which further dissociates and forms CO. 
Carbonyl groups are formed by oxygen addition to grain surfaces and are confirmed as intermediate species in the formation process. The yields of CO and CO$_2$ were found to be dependent on the thickness of the carbon layer. 
The astronomical relevance of the experimental findings is discussed. 

\end{abstract}

\keywords{Dust destruction (2268) --- Astrochemistry (75) --- Interstellar dust(836)
  --- Laboratory astrophysics(2004) --- Carbonaceous grains(201) --- Interstellar molecules(849)}


\section{Introduction} \label{sec:intro}

Solid CO and CO$_2$ are the main precursors for astrochemical reactions in the interstellar ice leading to complex organic molecules. The formed complex organic materials can be finally incorporated into planetesimals and cometesimals \citep{Bergin14}. This represents an indispensable source of organic carbonaceous material in meteorites and comets. The presence of CO ice in grain mantles is generally explained in terms of the direct condensation of CO molecules from the gas phase.

In the cold interstellar medium (ISM), the formation of CO$_2$ can be explained by the oxidation of CO ice. In this process, oxygen atoms resulting from the dissociation of oxygen-bearing molecules such as O$_2$ or H$_2$O by VUV (Ly-$\alpha$) photons or ions can react with CO. However, \citet{Raut11} found that the reaction of CO and $\textit{thermal}$ O atoms is relatively inefficient at making CO$_2$ because oxygen reacts primarily with O atoms to form O$_2$ and O$_3$. Alternatively, CO$_2$ can form on grain surfaces following the reaction CO~+ OH \citep{Watanabe02,Ioppolo09,Oba10,Ioppolo13}.

\begin{table*}[t]\label{Table01}
 \caption{Summary of performed experiments in this work.} 
    \centering 
\begin{tabular}{ccccccccc}
    \hline
    \hline
& \multicolumn{2}{c}{a-$^{13}$C Dust} & \multicolumn{2}{c}{Ice} & \multicolumn{2}{c}{X-Ray Fluence} & \multicolumn{2}{c}{Absorbed X-Ray} \\
Exp. & No. & Thickness & & Column Density & Photon & Energy & Photon &  Energy \\
& & (nm) & & (molecule cm$^{-2}$) & (photon cm$^{-2}$) & (eV cm$^{-2}$) & (photon cm$^{-2}$) & (eV cm$^{-2}$) \\
\hline
1 & I & 100 & H$_{2}$O & 5.0$\times$10$^{17}$ & 8.0$\times$10$^{18}$ & 5.9$\times$10$^{21}$ & 1.0$\times$10$^{18}$ & 7.4$\times$10$^{20}$ \\
2 & & & & 5.0$\times$10$^{17}$ & 7.7$\times$10$^{18}$ & 5.6$\times$10$^{21}$ & 9.9$\times$10$^{17}$ & 7.1$\times$10$^{20}$ \\
3 & & & & 5.0$\times$10$^{17}$ & 7.6$\times$10$^{18}$ & 5.5$\times$10$^{21}$ & 9.8$\times$10$^{17}$ & 7.0$\times$10$^{20}$ \\
4 & II & 100 & O$_{2}$ & 5.0$\times$10$^{17}$ & 7.6$\times$10$^{18}$ &  & 9.7$\times$10$^{17}$ &  \\
5 & & & $^{18}$O$_{2}$ & 5.0$\times$10$^{17}$ & 7.9$\times$10$^{18}$ &  & 1.0$\times$10$^{18}$ &  \\
6 & III & 300 & O$_{2}$ & 5.0$\times$10$^{17}$ & 7.6$\times$10$^{18}$ &  & 9.8$\times$10$^{17}$ &  \\
\hline
\end{tabular}
\end{table*}

Carbonaceous and siliceous dust grains are considered as main interstellar dust components providing surfaces for the formation of interstellar ice \citep{Draine03}. Water ice is the first ice layer accumulated on the surface of dust grains. There is only a limited number of studies on the interaction of interstellar ice, in particular water, with grain surfaces and the resulting photon-induced erosion of carbon grains at the dust-ice interface. Ice deposition and simultaneous photon or ion irradiation are processes that can frequently happen, resulting in a reduction of carbon grains in the ISM.

The ion-induced erosion of carbon grains due to CO and CO$_2$ formation at the interface between carbonaceous grains and ice layers was studied by a few groups only. \citet{Mennella04,Mennella06} studied the influence of energetic processing, with 30~keV He$^+$ ions and VUV photons on the interface of H$_2$O ice and hydrogenated carbon grains. The experimental data demonstrated that a few percent of carbon atoms bound in dust grains converted to volatile species such as CO and CO$_2$ with comparable yields. A similar experiment on dust erosion performed by \citet{Raut12} pointed to a temperature dependence of the CO$_2$ yield, but the authors did not detect CO formation after a bombardment with 100~keV protons. \citet{Fulvio12} investigated the possible role of carbon dust processing by oxygen to explain the CO$_2$ formation after energetic VUV irradiation. The photolysis of O$_2$ ice on the C-dust samples leads to CO$_2$, implying the direct oxygenation of carbon atoms at the contacting surfaces. In \cite{Sabri15}, O$_2$ and H$_2$O ice-coated carbonaceous grains, produced from $^{13}$C graphite targets by gas-phase condensation, were bombarded with 200~keV protons. A rate of 1.1~$\times$ 10$^{-15}$~nm ion$^{-1}$ was measured for a layer of O$_2$ ice. The molecules were exclusively formed from the carbon grains.

The aim of the current study is to understand carbon grain erosion at the interface grain/H$_2$O ice by processing with energetic X-ray photons in cold protoplanetary disk environments. For this purpose, the carbon ice samples were exposed to X-rays ranging from 250 to 1250~eV. The erosion of carbon was monitored by IR spectroscopy showing the formation of CO and CO$_2$ and scanning electron microscopy.

\section{Experiment} \label{sec:exp}

The Interstellar Energetic-Process System (IEPS), an ultrahigh vacuum (UHV) chamber with a base pressure of 5~$\times$ 10$^{-10}$~torr equipped with a cryostat, was used for the experiments on the erosion of carbon/ice layers. The setup is described in detail in \citet{Huang20}. Infrared spectra of the dust/ice samples were recorded using a mid-infrared Fourier-transform infrared spectrometer (FTIR) equipped with a mercury-cadmium-telluride infrared detector. 
H$_2$O (Merck, LC-MS grade), O$_{2}$ (Matheson, 99.99\%), or $^{18}$O$_{2}$ (Cambridge Isotope Laboratories, 95\%) ice was accumulated with a column density of $\sim$5~$\times$ 10$^{17}$ molecule cm$^{-2}$ on carbonaceous grains at a temperature of 13~K. The carbon grains were previously deposited on KBr substrates. The column density of the relevant molecules was estimated from the integration of IR absorbance peaks of the species using the modified Beer-Lambert Law (see equation (1) in \citealt{Huang20}). The applied band strength values are 2~$\times$ 10$^{-16}$, 7.6~$\times$ 10$^{-17}$, 7.8~$\times$ 10$^{-17}$, 1.1~$\times$ 10$^{-17}$, and 1.3~$\times$ 10$^{-17}$ cm molecule$^{-1}$ for H$_2$O ($\sim$3300 cm$^{-1}$), CO$_2$ ($\sim$2345 cm$^{-1}$), $^{13}$CO$_2$ ($\sim$2278 cm$^{-1}$), CO ($\sim$2142 cm$^{-1}$1), and $^{13}$CO ($\sim$2093 cm$^{-1}$), respectively \citep{Yamada1964, Jiang1975, Gerakines1995}. We know that a new band strength of CO$_2$ (i.e., 1.6~$\times$ 10$^{-16}$ cm molecule$^{-1}$) has been reported by \cite{Gerakines2015}. However, due to the lack of a new band strength for $^{13}$CO$_2$, we retain the original values reported in \cite{Gerakines1995} for both isotopes. All values are borrowed from the above literature on pure ice with some assumptions on its density and refractive index, where most of the uncertainties originate. The column density reported in this work can be calibrated when a better (more accurate) value is available. The initial formation rates were derived from a linear fit to the data within a fluence of 2~$\times$~10$^{18}$ photon cm$^{-2}$, and the formation kinetics (cross section) were derived from an exponential fit accompanied by a standard deviation as error bars.

Layers of amorphous 13-carbon grains (a-$^{13}$C dust hereafter) were produced by pulsed laser ablation of $^{13}$C graphite targets and subsequent condensation in a quenching gas atmosphere of helium. A scheme of the experimental setup and details are provided in \citet{Jaeger08}. A brief description of the a-$^{13}$C synthesis is provided in the following, and the possible contamination is discussed in Appendix~\ref{appendix_A}.
A pulsed Nd:YAG laser with a wavelength of 532~nm (second harmonic) was used to evaporate carbon from the rotating target. With laser pulses of 5~ns and pulse energies between 100 and 230~mJ, power densities between 9~$\times$ 10$^9$ and 7~$\times$ 10$^{10}$~W~cm$^{-2}$ were applied, generating temperatures of more than 4000~K in the condensation zone. The condensation of carbon particles is caused by collisions between the evaporated carbon atoms and clusters and further cooling by collisions with He atoms at a pressure of 4~torr in the ablation plume. The particles were extracted from the condensation zone through a nozzle and skimmer to form a freely propagating beam, which was finally deposited on 2~mm thick KBr substrates (Korth Kristalle GmbH) placed in the deposition chamber. A quartz microbalance was used to measure the thickness of the grain film. The deposition chamber is connected to an FTIR spectrometer that permits us to record the IR spectra of the deposited films in situ. 

The ice samples were irradiated by a soft X-ray beam from the beamline BL08B at the National Synchrotron Radiation Research Center (NSRRC, Taiwan). The X-ray spectrum covered photon energies from 250 to 1250 eV, with a flux of $\sim$2.1~$\times$ 10$^{15}$ photon cm$^{-2}$ s$^{-1}$ based on the size of the X-ray spot, which was 4~$\times$ 2~mm$^{2}$ \citep{Ciaravella19}. The incident and absorbed X-ray fluxes are reported in Appendix~\ref{appendix_B}. The irradiation was performed in steps with a total irradiation time of 60 minutes. Infrared spectra were taken with a resolution of 2~cm$^{-1}$. The performed experiments are summarized in Table 1.  At the end of the irradiation, the ice was heated to room temperature at a rate of 2~K~minute$^{-1}$.

Field emission scanning electron microscopy (FESEM) with a Zeiss (LEO) 1530 Gemini microscope with a maximum resolution of 2~nm was used to provide information on the morphology of the carbon layers before and after processing.  

\begin{figure*}[t]
	\begin{center}
		\includegraphics[width=\textwidth]{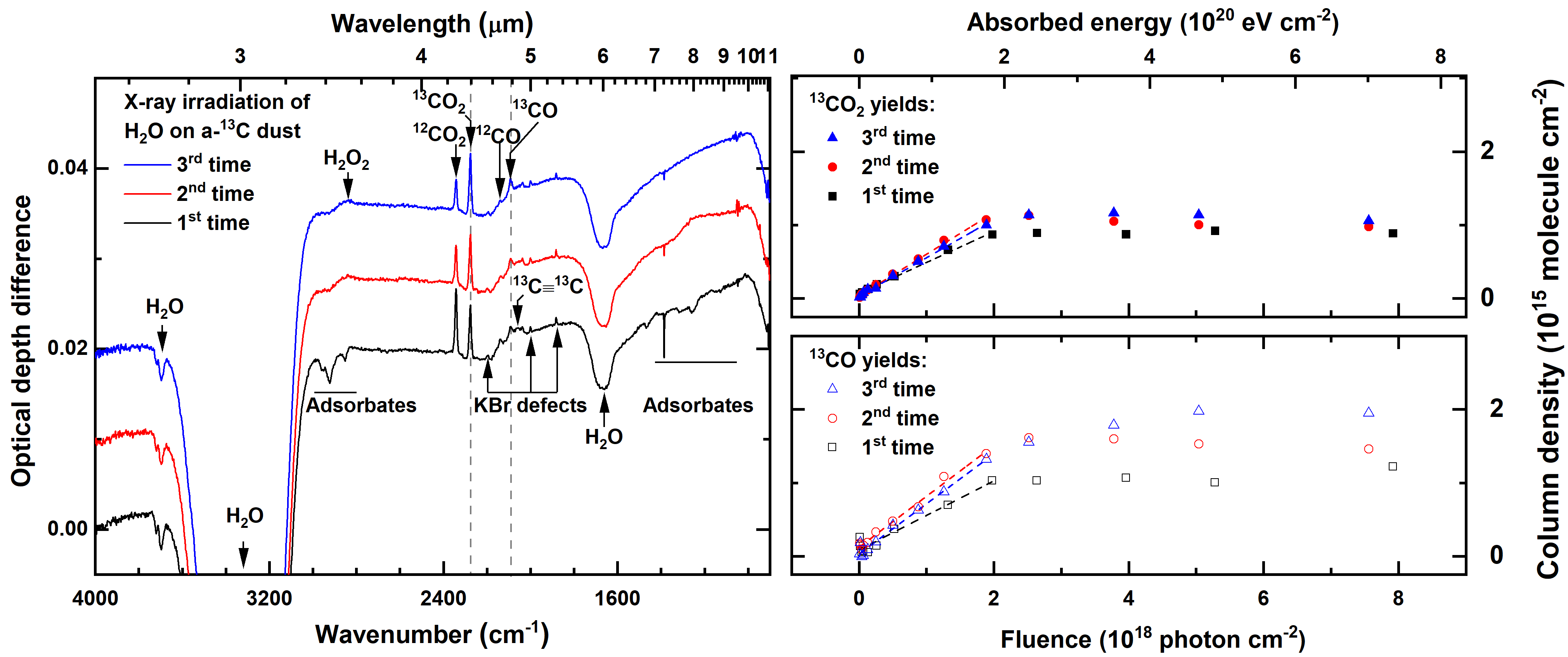}
		\caption{Left: the IR difference spectra obtained before and after X-ray irradiation of H$_2$O ice on a-$^{13}$C dust analog at 15~K for 60 minutes. The thickness of a-$^{13}$C dust is $\sim$100~nm, and the deposited abundance of H$_2$O is $\sim$5~$\times$ 10$^{17}$ molecule cm$^{-2}$. The dashed lines highlight the peak position of $^{13}$C-labeled products. The spectra are offset for clarity. Right: abundance evolution of the product $^{13}$CO$_2$ (solid) and $^{13}$CO (open) obtained from the X-ray irradiation of H$_2$O ice on the same a-$^{13}$C dust over X-ray photons fluence of (7.6--8.0)~$\times$ 10$^{18}$ photon cm$^{-2}$ (i.e., (7.0--7.4)~$\times$ 10$^{20}$~eV~cm$^{-2}$). The dashed lines show the linear fit for the initial formation rate.}
		\label{Fig1}
	\end{center}
\end{figure*}

\section{Results and Discussion}\label{sec:results}
\subsection{X-Ray Irradiation of \texorpdfstring{H$_2$O}{} Ice on \texorpdfstring{a-$^{13}$C}{} Dust}\label{subsec3.1}

The IR difference spectra obtained after X-ray irradiation of H$_2$O ice ($\sim$5~$\times$ 10$^{17}$ molecule cm$^{-2}$) deposited on a-$^{13}$C dust analog at $\sim$13~K for 60 minutes are shown in the left panel of Figure~\ref{Fig1}. Three irradiation experiments were repeated sequentially with the freshly deposited H$_2$O ice using the same a-$^{13}$C dust. The  H$_2$O features at $\sim$3300~cm$^{-1}$ (OH stretching) and $\sim$1670~cm$^{-1}$ (OH bending) 
are presented in negative bands along with peaks at 2960/2922/2850 cm$^{-1}$ (CH stretching) and 1467/1386/1317/1260 cm$^{-1}$ (CH bending) due to an organic adsorbate layer that is immediately formed under exposure to air (see Appendix~\ref{appendix_A}). These bands are most pronounced in the first experiment. However, the peak intensities of the contamination became smaller in the second and third experiments, implying that X-ray irradiation minimizes impurities. Besides the H$_2$O depletion, the IR feature of H$_2$O$_{2}$ is found at $\sim$2850~cm$^{-1}$, which is formed through the recombination of two OH radicals after H$_2$O dissociation. In addition, two strong peaks at 2342 and 2280~cm$^{-1}$ assigned to $^{12}$CO$_{2}$ and $^{13}$CO$_2$, respectively, appeared along with tiny peaks of $^{12}$CO (2137~cm$^{-1}$) and $^{13}$CO (2092~cm$^{-1}$) \citep{Sabri15}. The $^{13}$C-labeled products are the direct evidence of the interactions between H$_2$O ice and a-$^{13}$C dust, while $^{12}$C-labeled species are only considered as contributions of adsorbates to the product inventory. 

\begin{figure*}[t]
	\begin{center}
		\includegraphics[width=\textwidth]{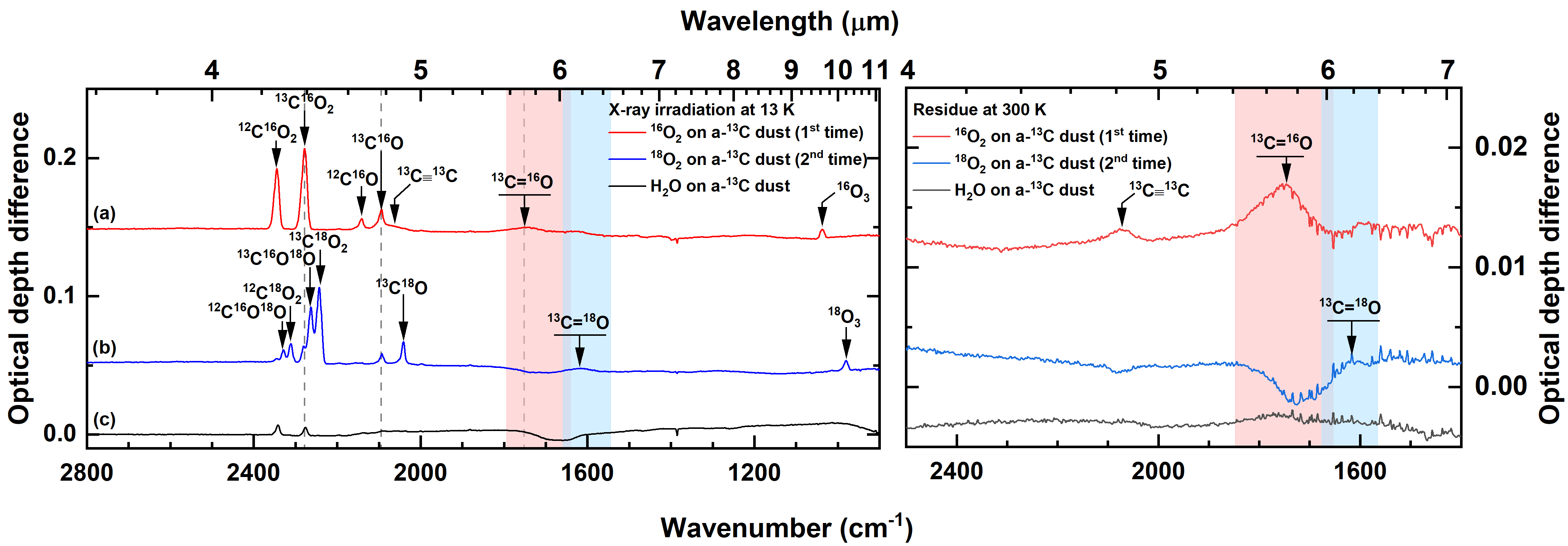}
            \caption{Left: the IR difference spectra obtained before and after X-ray irradiation of (a) $^{(16)}$O$_{2}$ and (b) $^{18}$O$_{2}$ ices on the same a-$^{13}$C dust analog as well as (c) H$_2$O on a-$^{13}$C dust analog at 15~K for 60 minutes. Right: the corresponding residual spectra measured at 300~K. The thickness of a-$^{13}$C dust is $\sim$100~nm, and the deposited abundances of H$_2$O and $^{(16/18)}$O$_{2}$ are $\sim$5~$\times$ 10$^{17}$ molecule cm$^{-2}$. The total fluence of X-ray photons is (7.6--7.9)~$\times$ 10$^{18}$ photon cm$^{-2}$. The dashed lines highlight the peak position of $^{13}$C-labeled products, and the shadowed area highlights the regions of the carbonyl group feature. The spectra are offset for clarity.
            }
            \label{Fig2}
	\end{center}
\end{figure*}

The formation of products, including $^{12}$C and $^{13}$C-labeled species, is monitored as a function of the incident photon fluence or absorbed energy fluence, showing that $^{12}$C and $^{13}$C-bearing products are formed simultaneously upon X-ray irradiation. In the case of carbon dioxide (see Fig. \ref{FigA3} in Appendix \ref{appendix_C}), the $^{12}$CO$_2$ yield is slightly higher than $^{13}$CO$_2$ by $\sim$50\% at the very beginning of irradiation (i.e., fluence $<$5~$\times$~10$^{17}$ photon cm$^{-2}$) in the first experiment. Moreover, the abundance (or formation efficiency) of $^{12}$CO$_2$ decreases in the later experiments, which is consistent with less adsorption of the dust sample in the second and third experiments. In contrast, $^{13}$CO$_2$ abundance (or formation efficiency) remains relatively constant. This suggests that the formation of $^{13}$C-bearing products is less affected by the surface contamination in the currently studied X-ray-driven ice-dust chemistry. The right panel of Figure~\ref{Fig1} reports the kinetic evolution of newly formed $^{13}$CO$_2$ (upper panel) and $^{13}$CO (bottom panel) monitored for the three sequential experiments using the same a-$^{13}$C dust analog as a function of photon fluence. The absolute yields of $^{13}$CO$_2$ have very similar formation curves and reach a limit at $\sim$1~$\times$ 10$^{15}$ molecule cm$^{-2}$. Given a nearly linear increase of the $^{13}$CO$_2$ yield at the beginning of irradiation (i.e., fluence $<$2~$\times$ 10$^{18}$ photon cm$^{-2}$), a zero-order approximation is applied to derive an “initial” formation rate; the $^{13}$CO$_2$ formation rates are (4.7~$\pm$~0.3)~$\times$ 10$^{-4}$, (6.0~$\pm$~0.1)~$\times$ 10$^{-4}$, and (5.5~$\pm$~0.1)~$\times$ 10$^{-4}$ molecule photon$^{-1}$ for the first, second, and third experiments, respectively. In order to minimize the influence of the adsorbate layer, we report an averaged value of (5.7~$\pm$~0.2)~$\times$ 10$^{-4}$ molecule photon$^{-1}$ from the last two experiments. Since not all X-ray photons are absorbed by H$_2$O ice, it is ideal for reporting the product evolutions as a function of the absorbed energy fluence (as shown on the upper axis in Fig.~\ref{Fig1}). The formation rate is derived directly from a linear fit to the kinetic evolution as a function of the absorbed energy fluence. The fitting results give a universal formation rate in units of molecule per absorbed energy, i.e., an averaged value of (5.9~$\pm$ 0.2)~$\times$ 10$^{-6}$ molecule eV$^{-1}$ from the last two experiments, which can be directly compared with other (future) energetic processing. The estimated absorbed energy is only considered as an approximation (upper limit), given that not all energy will contribute to interfacial reactions. The detailed calculation of the incident photon and absorbed energy fluence is available in Appendix \ref{appendix_B}. The $^{13}$CO formation shares the “initial” forming rate in three experiments; an averaged rate of (7.5~$\pm$ 0.4)~$\times$ 10$^{-4}$ molecule photon$^{-1}$ (i.e., (7.4~$\pm$ 0.4)~$\times$ 10$^{-6}$ molecule eV$^{-1}$) was derived from two last experiments, but a larger scatter was noticed of the saturation abundance varying from 1.2~$\times$ 10$^{15}$ to 2.0~$\times$ 10$^{15}$ molecule cm$^{-2}$. Although higher product yields were found in later experiments (probably due to fewer adsorbates occupying the reaction sites), the same initial formation rates imply that these products were formed under identical conditions in each experiment. The O atoms from H$_2$O dissociation are the only species contributing to the product formation, given that no hydrogenated species were found. Therefore, the maximum yield is limited by the available $^{13}$C sites in the dust analog and other competitive reactions consuming O atoms. The chemical relation between $^{13}$CO and $^{13}$CO$_2$ remains unclear, given that their formation rate is too close to distinguish the order of the product’s generation firmly. This is partially due to the fact that carbon monoxide and carbon dioxide are known to be chemically linked through multiple reaction channels such as CO~+ O~$\rightarrow$ CO$_{2}$, CO~+ CO*~$\rightarrow$ CO$_{2}$~+ C, CO~+ OH~$\rightarrow$ HOCO~$\rightarrow$ CO$_{2}$~+ H, and CO$_2$~$\xrightarrow{hv}$ CO~+ O in the solid state under different interstellar environments \citep{Watanabe07,Ioppolo11,Chen14}. 


\subsection{Surface Reaction Mechanism}\label{subsec3.2}

To pinpoint the possible product formation route, two experiments using different isotope-labeled oxygen ice, e.g., $^{16/18}$O$_{2}$, were sequentially performed using the same a-$^{13}$C dust analog under identical conditions. The thickness of a-$^{13}$C dust is $\sim$100 nm, and the initial abundances of oxygen ice ($^{(16)}$O$_2$ and $^{18}$O$_2$) are expected to be $\sim$5$\times$10$^{17}$ molecule cm$^{-2}$, following the identical deposition setting as H$_2$O. In the following, the common isotope of oxygen atoms will be labeled as $^{(16)}$O. Figure~\ref{Fig2} shows the IR difference spectra obtained after X-ray irradiation of $^{(16)}$O$_{2}$ (a) and $^{18}$O$_{2}$ (b) ices at 13~K for 60 minutes and the corresponding IR spectra of the residues after warming up to 300~K. The IR spectrum (c) of X-ray irradiated H$_2$O ice on a-$^{13}$C is presented for comparison. In the spectrum (a) of Figure~\ref{Fig2}, $^{13}$C$^{(16)}$O$_{2}$ and $^{13}$C$^{(16)}$O are observed at 2280 and 2092~cm$^{-1}$, respectively, accompanied by $^{12}$C$^{(16)}$O$_{2}$ and $^{12}$C$^{(16)}$O resulting from adsorbate contributions. In addition, ozone ($^{(16)}$O$_3$) originating from the radiolysis of O$_2$ can be observed at $\sim$1038~cm$^{-1}$ . 
In contrast to the experiment with H$_2$O ice, a feature of the carbonyl functional group $^{13}$C$\dbond$$^{(16)}$O becomes visible at $\sim$1750~cm$^{-1}$, given no OH bending mode is present. Moreover, such a broad IR profile implies that this feature ranging from 1660 to 1840~cm$^{-1}$ is not simply due to a single molecular absorption but a broad distribution of $^{13}$C$\dbond$$^{(16)}$O groups on the surface of a-$^{13}$C dust, which became part of the refractory material. The residual spectrum at 300~K also confirms the appearance of $^{(16)}$O bound to the a-$^{13}$C dust. In the spectrum (b) of Figure~\ref{Fig2} using $^{18}$O$_{2}$ ice deposited on the previously $^{(16)}$O-attached dust, the IR spectrum shows a series of isotope-labeled carbon dioxides ($^{13}$C$^{18}$O$_{2}$, $^{13}$C$^{(16)}$O$^{18}$O, and $^{13}$C$^{(16)}$O$_{2}$ at 2280, 2263, and 2243~cm$^{-1}$, respectively) and carbon monoxides ($^{13}$C$^{18}$O and $^{13}$C$^{(16)}$O at 2092 and 2041 cm$^{-1}$, respectively) upon X-ray irradiation. Moreover, the previously formed refractory $^{13}$C$\dbond$$^{(16)}$O peak decreased, while the peak of $^{13}$C$\dbond$$^{18}$O appeared at $\sim$1610~cm$^{-1}$ and remained after warming up to 300~K. It is also noted that the IR peak intensities of products detected in O$_{2}$ ice are $>$10 times higher than those in H$_{2}$O ice since there is no competitive reaction involving  H$_{2}$O fragments.

\begin{figure}[t]
	\begin{center}
		\includegraphics[width=85mm]{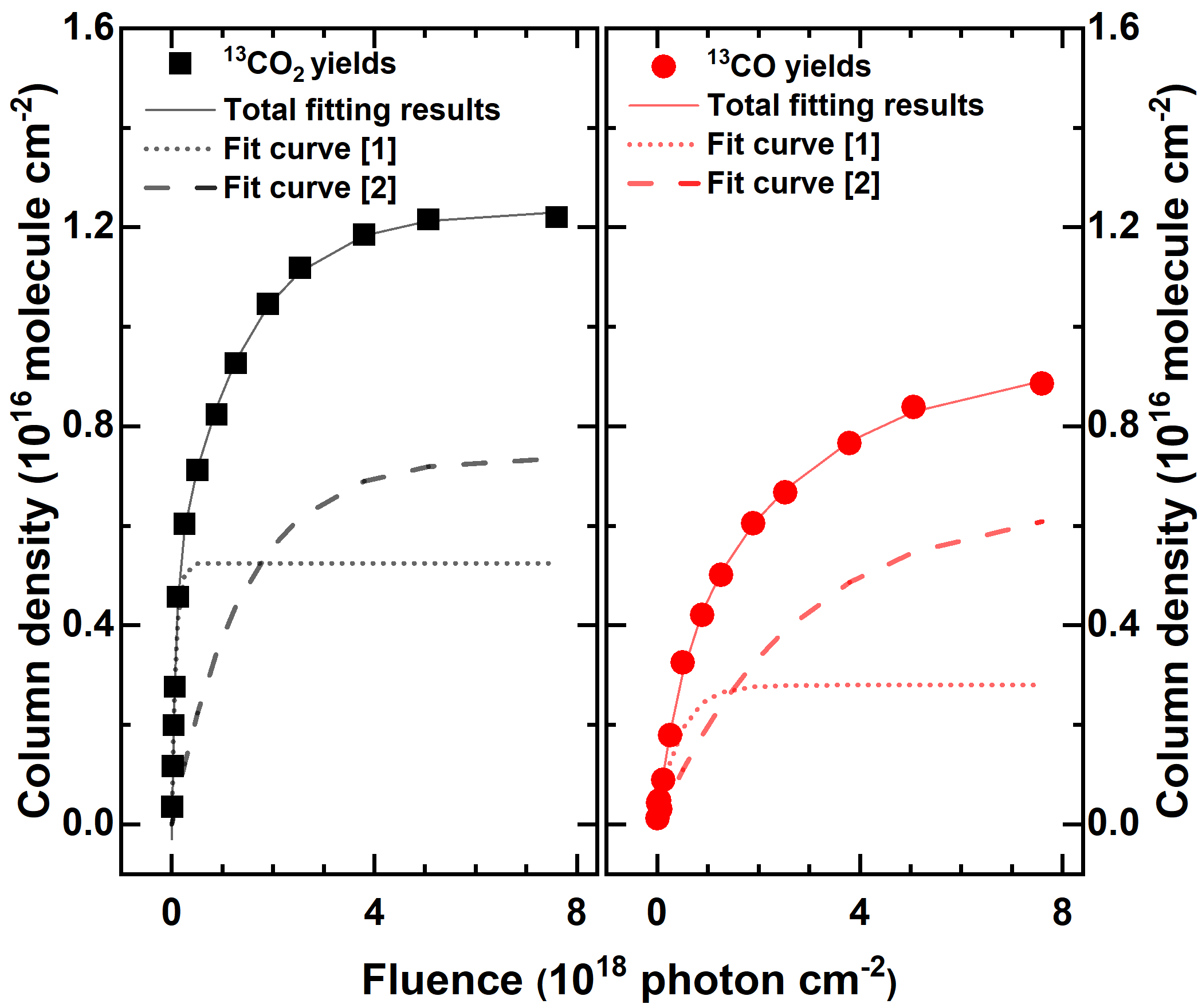}
		\caption{Abundance evolution of the product $^{13}$CO$_2$ (left) and $^{13}$CO (right) obtained from the X-ray irradiation of O$_{2}$ ice on a-$^{13}$C dust sample with a thickness of 100~nm over photon fluence of 7.6~$\times$ 10$^{-18}$ photon cm$^{-2}$ 
        The solid lines show the total fitting results and the dotted and dashed represent the fitting curves [1] and [2], respectively.}
		\label{Fig3}
	\end{center}
\end{figure}

The experimental results imply that the carbonyl group formation is an intermediate step contributing to the product formation. The X-ray-induced reaction mechanisms forming the carbonyl group are discussed in the following. The bulk ice chemistry triggered by X-ray photons has been experimentally investigated on interstellar relevant molecules, including H$_2$O, CO, and NH$_3$ \citep{MunozCaro19}. It is well accepted that X-ray photons mainly interact at the atomic level rather than molecular bonds; the K-edge absorption of O atoms and C atoms takes place at $\sim$532 and $\sim$284~eV, respectively, triggering Auger electrons in addition to primary photoelectrons. These newly generated electrons are responsible for the following cascade reactions, including excitation and ionization, leading to the formation of secondary electrons until they lose their energy.  In this work, the majority of X-ray photons is assumed to be absorbed by O atoms from 540 to 1000~eV since the incident photon flux at $\sim$290~eV is one order of magnitude lower than the maximum intensity (see Figure~\ref{FigA1}). The low-energy electrons further break O$_2$ or H$_2$O through excitation impacts or dissociative electron attachment (DEA), ultimately forming neutral/ionic atoms (or radicals). O atom addition reactions to the available reaction sites of a-$^{13}$C dust result in oxygenated carbon dust (i.e., $^{13}$C$\dbond$$^{(16/18)}$O group; \citealt{Backreedy02}).

There are two possible product reaction pathways relying on the newly formed carbonyl group. First, detachment of the carbonyl groups $^{13}$C$\dbond$$^{(16)}$O from a-$^{13}$C dust due to interactions with secondary electrons results in $^{13}$C$_n$$^{(16)}$O ($n$~$\ge$ 1) species. The newly released $^{13}$C$^{(16)}$O molecule can further associate with $^{(16/18)}$O atoms, forming $^{13}$C$^{(16/18)}$O$_{2}$. In this scenario, $^{13}$C$^{(16)}$O is expected to form efficiently before the production of $^{13}$C$^{16/18}$O$_2$. Second, sequential $^{18}$O-atom attachment to refractory $^{13}$C$\dbond$$^{(16)}$O carbonyl groups forms doubly oxygenated species. The excess reaction energy breaks up the C$\sbond$C bond to the amorphous carbon resulting in $^{13}$C$^{(16)}$O$^{18}$O release \citep{Shi15}. According to the results shown in Figure~\ref{Fig2}, the a-$^{13}$C dust surface is also chemically active. The continuous addition of $^{18}$O atoms triggers the transition from $^{13}$C$\dbond$$^{(16)}$O to $^{13}$C$\dbond$$^{18}$O bound to the dust analog, eventually enriching the amount of $^{13}$C$^{18}$O$_{2}$, which is securely observed in the spectrum (b). 

The kinetic evolution of the products $^{13}$C$^{(16)}$O$_{2}$ and $^{13}$C$^{(16)}$O obtained in the X-ray irradiation of the $^{(16)}$O$_{2}$ ice experiment over 60 minutes are presented in Figure~\ref{Fig3}. The $^{13}$C$^{(16)}$O$_{2}$ yield increases rapidly upon X-ray irradiation and stabilizes at $\sim$1.3~$\times$ 10$^{16}$ molecule cm$^{-2}$ after reaching a fluence of $\sim$4~$\times$ 10$^{-18}$ photon cm$^{-2}$. The formation curve of $^{13}$C$^{(16)}$O differs from that of $^{13}$C$^{(16)}$O$_{2}$; its absolute yield and formation efficiency are smaller than those of $^{13}$C$^{(16)}$O$_{2}$. The product ratio of $N$($^{13}$C$^{(16)}$O)/$N$($^{13}$C$^{(16)}$O$_{2}$) varies from $\sim$0.3 at the beginning to $\sim$0.7 at the end of the X-ray fluence, pointing to a later formation of $^{13}$C$^{(16)}$O than $^{13}$C$^{(16)}$O$_{2}$. A proper temporal fit to their formation curves is only achieved with at least two exponential terms, i.e., $N$(molecule)~= $\alpha_1(1-\mathrm{exp}(-\sigma_1\phi))+\alpha_2(1-\mathrm{exp}(-\sigma_2\phi))$, where $\alpha$ is the saturation (i.e., maximum yield when reaching the equilibrium state) in molecule cm$^{-2}$, $\sigma$ is the apparent formation cross section in cm$^2$, and $\phi$ is the photon fluence in photon cm$^{-2}$. In the formation process of $^{13}$C$^{(16)}$O$_{2}$, the derived $\alpha_1$ is (5.2~$\pm$ 0.2)~$\times$ 10$^{15}$ molecule cm$^{-2}$ with $\sigma_1$~= (1.3~$\pm$ 0.1)~$\times$ 10$^{-17}$~cm$^2$ and $\alpha_2$ is (7.4~$\pm$ 0.2)~$\times$ 10$^{15}$ molecule cm$^{-2}$ with $\sigma_2$~= (7.1~$\pm$ 0.4)~$\times$ 10$^{-19}$~cm$^2$. 

The derived kinetics (i.e., saturation and cross section) suggest two contribution formation scenarios. The different formation efficiencies could be related to carbonyl groups intermediately formed either on aliphatic or aromatic structural units that build up the amorphous carbon dust. This is in line with the observation of a broad IR peak profile in Figure~\ref{Fig2}. Alternatively, the excess energy after DEA could trigger O-atom diffusion into deeper/inner sites of fullerene-like nanoparticles \citep{Jaeger08}. Future study is highly needed to investigate the possible erosion sites in a-$^{13}$C dust. We cannot exclude the possibility of different kinds of photoelectron contributions to product formation.  

For $^{13}$C$^{(16)}$O, the corresponding saturation (i.e., $\alpha_1$~= (2.8~$\pm$ 0.9)~$\times$ 10$^{15}$ and $\alpha_2$~= (6.5~$\pm$ 0.7)~$\times$ 10$^{15}$ molecule cm$^{-2}$) and formation cross section (i.e., $\sigma_1$~= (2.3~$\pm$ 0.8)~$\times$ 10$^{-18}$ and $\sigma_2$~= (3.6~$\pm$ 0.8)~$\times$ 10$^{-19}$~cm$^2$) values are smaller than those of $^{13}$C$^{(16)}$O$_{2}$. Product kinetics suggests that the formation of carbon monoxide is generally later than carbon dioxide. The dominant $^{13}$C$^{(16)}$O$_{2}$ formation upon X-ray irradiation supports the conclusion that $^{13}$C$^{(16)}$O is most likely the consequence of $^{13}$C$^{(16)}$O$_{2}$ dissociation rather than a direct detachment of C$\dbond$O groups bound to the surface of grains. 

\begin{figure}[t]
	\begin{center}
		\includegraphics[width=85mm]{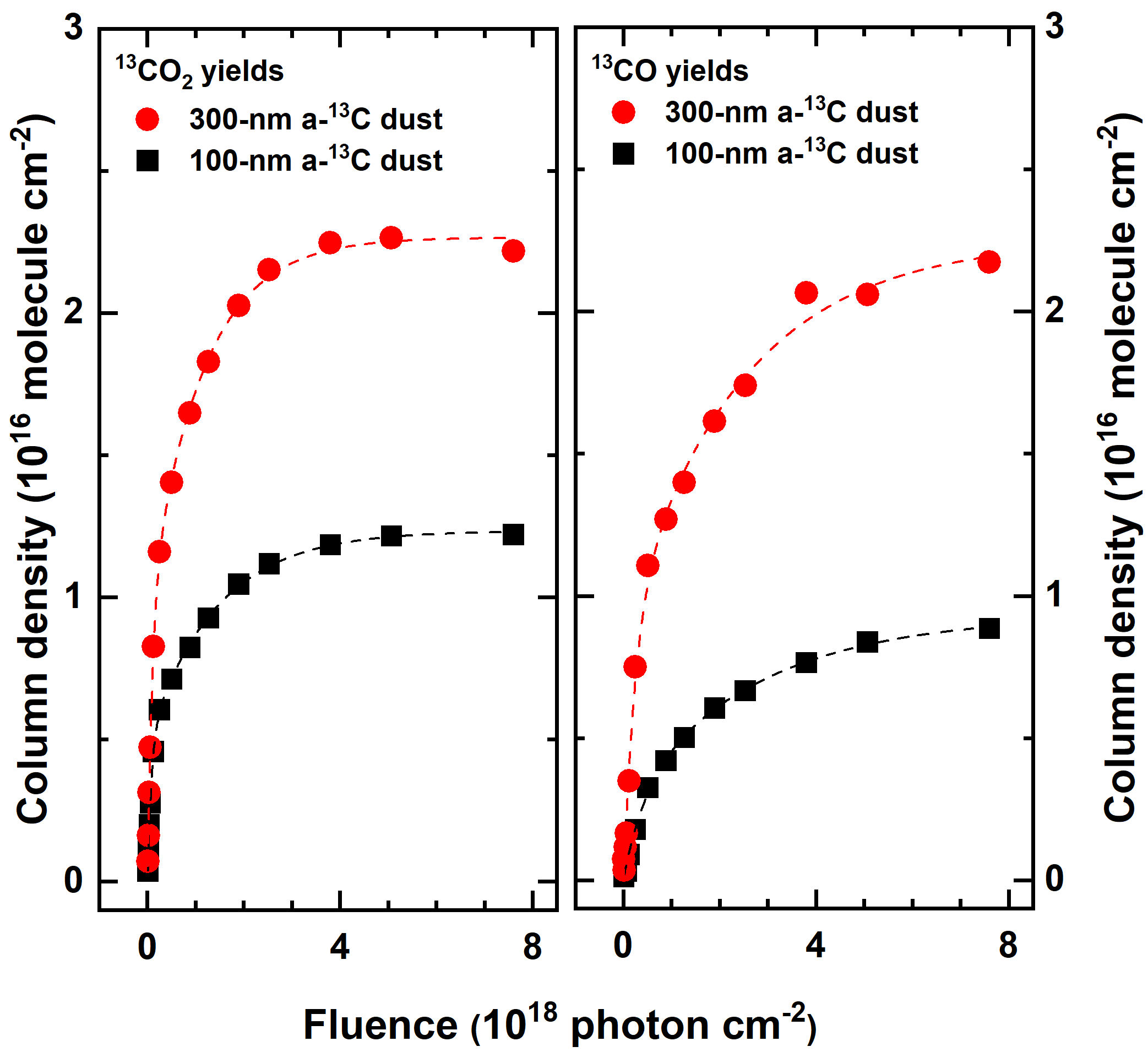}
		\caption{Abundance evolution of the products $^{13}$CO$_2$ (left) and $^{13}$CO (right) obtained from the X-ray irradiation of O$_{2}$ ice on two a-$^{13}$C dust samples with a thickness of $\sim$100 and $\sim$300~nm over photon fluence of (7.6--7.9)~$\times$ 10$^{18}$ photon cm$^{-2}$. 
        The dashed lines show the total fitting results.}
		\label{Fig4}
	\end{center}
\end{figure}

\subsection{Thickness-dependent \texorpdfstring{$^{13}$CO$_2$}{} Formation}\label{subsec3.3}

Figure~\ref{Fig4} shows a thickness-dependent formation of $^{13}$CO$_2$ and $^{13}$CO from the X-ray irradiation of O$_2$ ice on $\sim$100 and $\sim$300~nm a-$^{13}$C dust layers as a function of photon fluence. After reaching a fluence of $\sim$8~$\times$ 10$^{18}$ photon cm$^{-2}$, the absolute yields of $^{13}$CO$_2$ and $^{13}$CO on $\sim$300~nm a-$^{13}$C are $\sim$2.3~$\times$ 10$^{16}$ and $\sim$2.2~$\times$ 10$^{16}$ molecule cm$^{-2}$, respectively, which are 2--3 times higher than the yields from $\sim$100~nm grains. The ratio of $^{13}$CO over $^{13}$CO$_2$ varies from $\sim$0.2 to $\sim$1.0, confirming the formation of $^{13}$CO following the $^{13}$CO$_2$ dissociation, as reported in the measurement on $\sim$100~nm grains.

Similarly, the formation curves are at the best fit with two formation components. The derived apparent formation cross sections (i.e., $\sigma$) are (1.0~$\pm$ 0.1)~$\times$ 10$^{-17}$ and (9.0~$\pm$ 0.7)~$\times$ 10$^{-19}$~cm$^2$ for $^{13}$CO$_2$ and (4.3~$\pm$ 1.2)~$\times$ 10$^{-18}$ and (4.1~$\pm$ 1.1)~$\times$ 10$^{-19}$~cm$^2$ for $^{13}$CO. These values are in good agreement with those obtained with $\sim$100 nm dust (within the uncertainties), suggesting a similar formation mechanism on both layers but a higher yield on the thicker a-$^{13}$C sample. The experimental findings point out that the $\sim$300 nm dust layer provides a larger surface area and, therefore, more reaction sites for oxygen addition reactions. 

\subsection{Erosion of \texorpdfstring{a-$^{13}$C}{} Dust}\label{subsec3.4}
\begin{figure*}[t]
	\begin{center}
		\includegraphics[width=160mm]{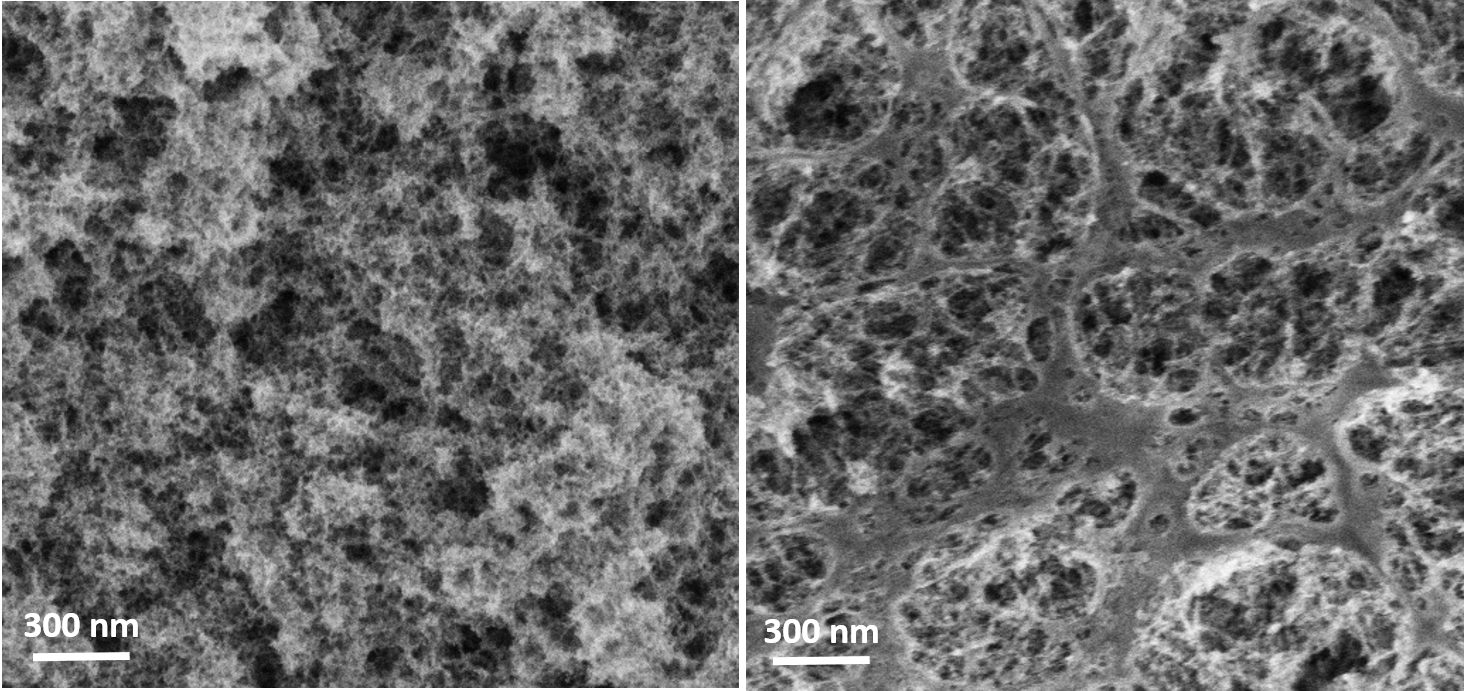}
		\caption{Scanning electron microscope images of the original carbon material (left) and after ice deposition and subsequent irradiation with X-rays (right). The original thickness of a-$^{13}$C dust is $\sim$300~nm. Left: the carbon grains are 3--5~nm in size and form fractal entities that finally produce a very porous, three-dimensional layer. Right: the erosion of carbon is clearly visible by a strong flattening of the carbon structure.}
		\label{Fig5}
	\end{center}
\end{figure*}

The erosion of carbon grains and recognizable changes in the morphology of eroded and non-eroded parts of the carbon sample can be monitored by using FESEM, which provides information on the topology of the sample (see Figure~\ref{Fig5}). The original carbon film before the deposition of ice layers and subsequent X-ray irradiation looks very fluffy with piles of aggregates of carbon grains.  After the deposition of water or O$_2$ ice and following irradiation with X-rays, distinct modifications of the morphology and topography of the samples are visible. The thickness of the carbon layers was significantly reduced. The flattening of the processed carbon film indicates the removal of the upper fluffy piles of grains. Only a reduced amount of carbon remained after irradiation.  
For the carbon sample shown in Figure~\ref{Fig5}, the amount of carbon material removed by the erosion was estimated to be at least 60\%  of the original thickness. The changes in the morphological characteristics indicate a strong penetration of water or oxygen molecules into the porous structure, leading to an efficient erosion process of carbon grains from all sides.

\section{Astrochemical Implications and Conclusion}\label{sec:implication and conclusion}

This work presents, for the first time,  the solid-state interface reactions between H$_2$O (or O$_2$) ice and amorphous carbonaceous dust grains induced by the impact of X-ray photons at 13~K. The experimental results firmly show the formation of carbon dioxide followed by the formation of carbon monoxide using isotope-labeled $^{13}$C dust samples. In a well-defined control experiment involving $^{16}$O$_2$ and $^{18}$O$_2$ ice, possible formation mechanisms were studied; sequential $^{16/18}$O-atom attachment reactions to active carbon sites on a-$^{13}$C dust result in the doubly isotope-labeled product C$^{16}$O$^{18}$O.

The protoplanetary disk is a stage before the formation of planetary systems. The chemical ingredients within the midplane, where planetesimals form, might control the final composition. Moreover, H$_2$O ice, the most abundant interstellar solid-state molecule likely inherited from molecular (translucent) clouds, has been suggested to facilitate rapid coagulation of dust grains \citep{Okuzumi12}. Astrochemical models of protoplanetary disks have cataloged three different water chemistry regimes: (1) gaseous water within the ice line of 160~K, (2) cold water located beyond the ice line in the outer disk, and (3) warm water vapor synthesized in the surface layers of disks (see Figure~25 in a review paper by \citealt{vanDishoeck21}). Young stellar objects typically emit soft X-rays ($E$~$\leq$ 1~keV) due to the stellar coronal activities \citep{Gudel09}. Class I/II protostars are generally considered to be bright X-ray sources with luminosities of about 10$^{28}$--10$^{31}$~erg~s$^{-1}$ \citep{Imanishi01}. Such intense X-rays can easily penetrate the thick inner disk, reach the outer midplane, and trigger surface-grain chemistry, ultimately altering the chemical composition of gas and ice \citep{Almeida14, Walsh14, Jimenez18, Ciaravella19}. 

This experimental work studying the X-ray-induced interactions between H$_2$O ice and a-$^{13}$C dust for an absorbed energy fluence of $\sim$7.0~$\times$ 10$^{20}$~eV~cm$^{-2}$ is well situated in the aforementioned “regime 2” where H$_2$O ice remains on dust grains and receives X-ray photons emitted by the central protostar. According to the astrochemical 2D protoplanetary model described by \citet{Walsh14}, the X-ray flux obtained in the midplane 
is $\sim$2~$\times$ (10$^{-7}$--10$^{-5}$)~erg~cm$^{-2}$~s$^{-1}$, i.e., 1~$\times$ (10$^{5}$--10$^{7}$)~eV~cm$^{-2}$~s$^{-1}$. It leads to the total energy fluence of 4~$\times$ (10$^{18}$--10$^{21}$)~eV~cm$^{-2}$, given a typical protoplanetary lifetime of 10$^{6}$--10$^{7}$~yr. The proposed formation mechanism is likely to occur at the contacting surfaces of carbonaceous dust grains in this regime. The vertical mixing of icy grains from midplane to the upper layers of protoplanetary disks suggests that the studied mechanism is also relevant to levitated surface layers or even enhanced due to a higher X-ray flux available  \citep{Semenov2011, Furuya2013}. Moreover, H$_2$O-coated grains can exist in a relatively wide temperature range ($\leq$150 K). Depending on the dust temperature, the newly formed CO$_2$ might remain in the ice or sublimate into the gas phase when the dust temperature is beyond its desorption threshold. More experiments are desired to explore the vertical contribution as a function of the temperature gradient along the disk height. The obtained thickness-dependent yield of CO$_2$ suggests that X-ray-induced O-atom reactions with C dust are feasible when icy dust grains start coagulating to form icy pebbles. It is important to note that different types of ice-dust compositions exist in protoplanetary disks, and their CO$_2$ production efficiency might vary along different physicochemical parameters. Therefore, the derived production rates/cross sections of C-bearing species in this study are rigorously reported for laboratory comparison.

In conclusion, the main experimental findings are summarized below:
\begin{itemize}
  \item The interface reactions between H$_2$O and a-$^{13}$C dust induced by X-ray photons (250--1250~eV) at $\sim$13~K result in the erosion of carbon grains and the production of $^{13}$CO$_2$ and $^{13}$CO. The initial formation rates were derived to be (5.9~$\pm$ 0.2)~$\times$ 10$^{-6}$ molecule eV$^{-1}$ and (7.4~$\pm$ 0.4)~$\times$ 10$^{-6}$ molecule eV$^{-1}$ for $^{13}$CO$_2$ and $^{13}$CO, respectively.  
  \item The mechanism of product formation is verified by using isotope-labeled $^{16/18}$O atoms; the successive O-atom addition to carbonyl groups on the surface of grains forms doubly oxygenated species, which are further released as intact $^{13}$CO$_2$ probably due to excess reaction energy. Kinetic analysis suggests that $^{13}$CO is a second-generation product and comes most likely from the $^{13}$CO$_2$ dissociation.
  \item The $^{13}$CO$_2$ yield is dependent on the thickness of a-$^{13}$C dust grains, probably due to a larger ice-dust contacting area. Experimental results imply that in the cold midplane of protoplanetary disks, where X-ray processing dominates the solid-state chemistry, the studied O-atom addition reactions play an essential role in enriching the content of carbon-bearing species besides gas accretion.
\end{itemize}


\section*{Acknowledgments}

We acknowledge the NSRRC general staff for running the synchrotron radiation facility. We also thank Dr. A. Chainani, the spokesperson of BL08B in NSRRC. The research is funded by the Deutsche Forschungsgemeinschaft with grant JA 2107/10-1 (project No. 468269691).
C.J. also acknowledges the support by the Research Unit FOR 2285 “Debris Disks in Planetary Systems” of the Deutsche Forschungsgemeinschaft through grant JA 2107/3-2 (project No. 262443618). K.J.C. is grateful for support from the Danish National Research Foundation through the Center of Excellence “InterCat” (Grant agreement no.: DNRF150) and the Dutch Research Council (NWO) via a VENI fellowship (VI.Veni.212.296). This work has also benefited from financial support from National Science and Technology Council, Taiwan, under grant Nos. NSTC 110-2628-M-008-004-MY4 and NSTC 110-2923-M-008-004-MY3 (Y.-J.C.). We thank G. Rouillé for stimulating discussions.

\appendix

\restartappendixnumbering

\section{Synthesis of \texorpdfstring{a-$^{13}$C}{} Dust}\label{appendix_A}

Figure~\ref{FigA2} shows the IR spectra of the amorphous carbonaceous dust analogs labeled with isotope $^{13}$C (a-$^{13}$C dust hereafter) after production and exposure to air. The newly produced a-$^{13}$C dust has a relatively structureless spectrum with a broad plateau in the range of 1000--1600~cm$^{-1}$ suggesting that the refractory material mainly consists of aromatic and fullerene-like structures containing sp$^2$-hybridized carbon \citep{Jaeger08}. Once a-$^{13}$C dust is exposed to air, an adsorbate layer is formed, unavoidably presenting several IR absorption features such as OH (CH) stretching and bending, CO stretching, and other unidentified contamination peaks. These contaminations are largely depleted upon either annealing at elevated temperatures ($\sim$500~K) or X-ray irradiation (in this work), as shown in Figure~\ref{Fig1}. Nevertheless, adsorbates should be rich in the common carbon isotope $^{12}$C due to the natural $^{13}$C/$^{12}$C ratio of $\sim$0.01. Therefore, the contribution originating from adsorbates and a-$^{13}$C dust is distinguishable through the isotope-labeled products. Minor peaks at 1882, 2000, 2036, and 2194~cm$^{-1}$ have been identified in a control irradiation experiment of bare a-C dust, and they are due to defects of KBr.

\setcounter{figure}{0}
\begin{figure}[b]
	\begin{center}
		\includegraphics[width=85mm]{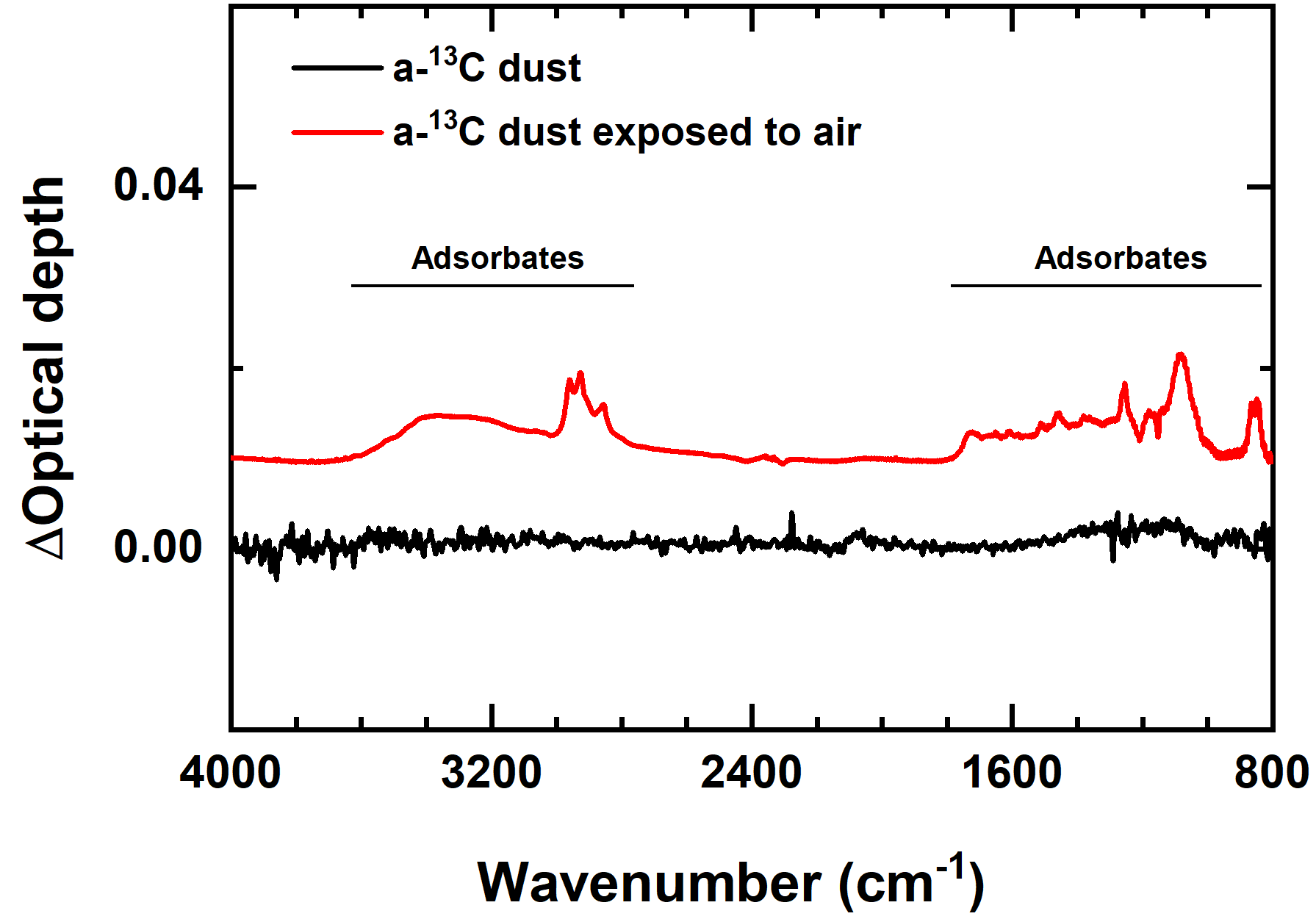}
		\caption{IR spectra of a-$^{13}$C obtained after deposition and exposure to air. The spectra are offset for clarity.}
		\label{FigA2}
	\end{center}
\end{figure}

\section{X-Ray Flux}\label{appendix_B}

\begin{figure}[t]
	\begin{center}
		\includegraphics[width=85mm]{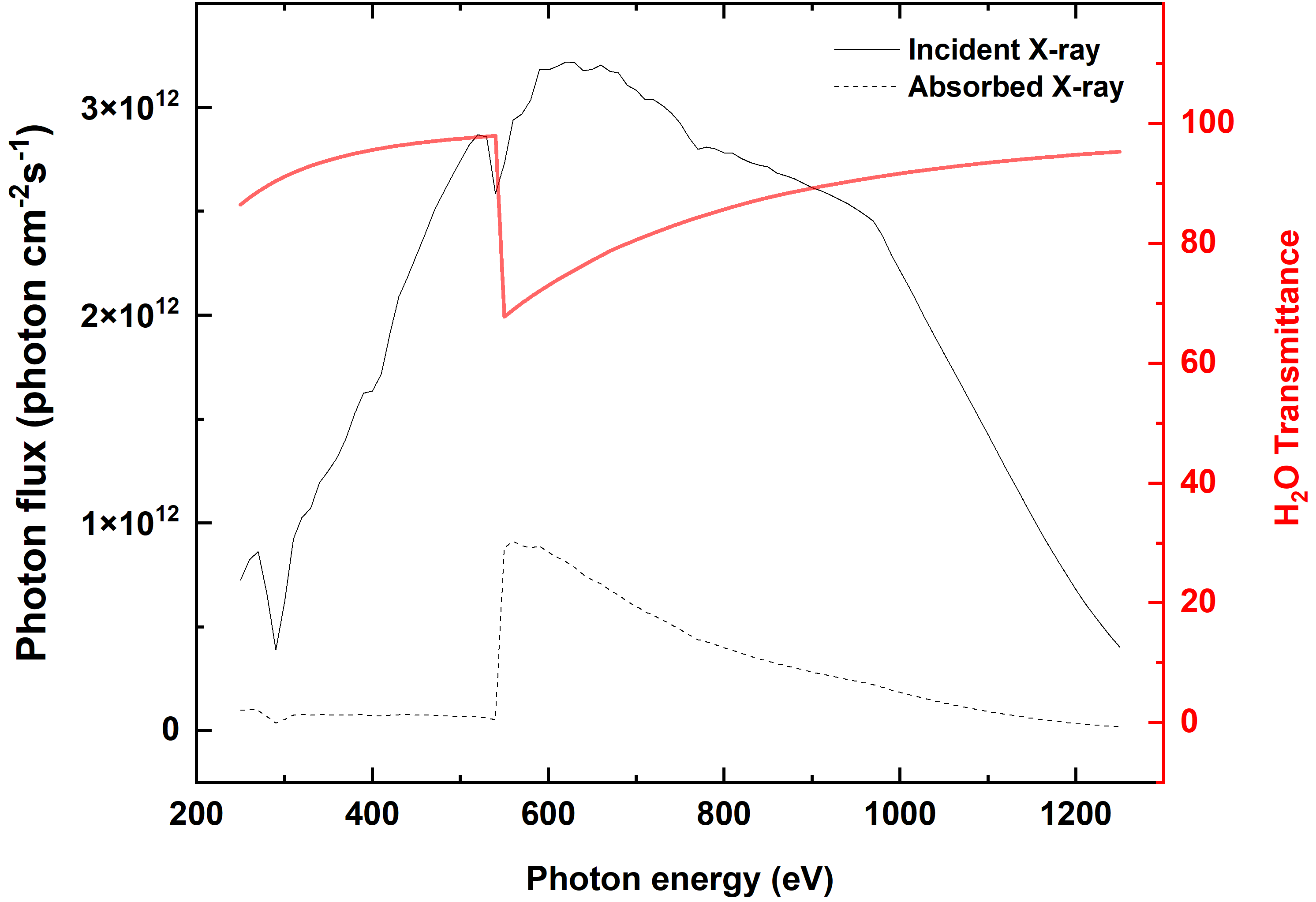}
		\caption{The X-ray photon flux and H$_2$O transmittance as a function of photon energy.}
		\label{FigA1}
	\end{center}
\end{figure}

\begin{figure}[t]
	\begin{center}
		\includegraphics[width=100mm]{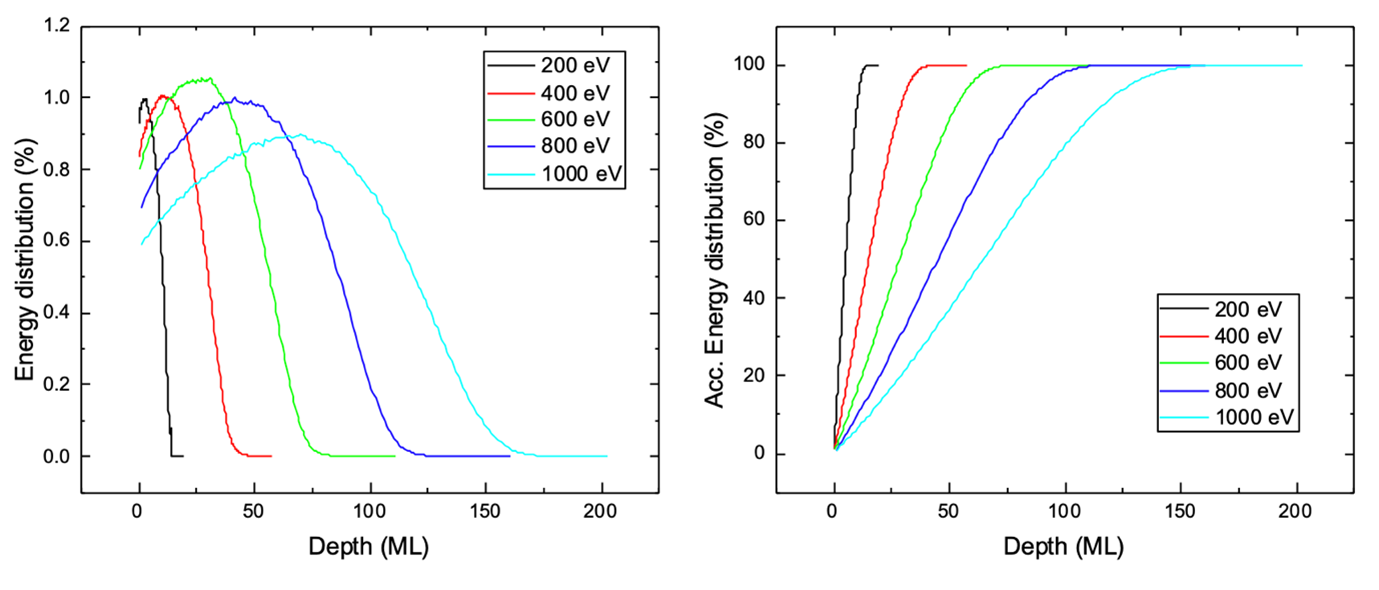}
		\caption{CASINO simulation for the energy deposition distribution (left panel) and accumulated deposited energy distribution (right panel) for electron energies of 200, 400, 600, 800, and 1000 eV as a function of penetration depth. The applied H$_2$O density is 0.87 g cm$^{-3}$; 1 nm = 2.91 ML.}
		\label{FigA4}
	\end{center}
\end{figure}

Figure~\ref{FigA1} presents spectral characteristics of the X-ray flux in the experiments, including the incident photon flux and the photon flux absorbed by a mantle of H$_2$O ice. The X-ray photon flux was monitored using an in-line nickel mesh (with about 90\% optical transmission) that was calibrated using a traceable photodiode from International Radiation Detectors, Inc. As the X-rays propagated inside the ice, their flux decreased exponentially, and the X-Ray Database\footnote{https://henke.lbl.gov/} was used to compute this decrease, which included photoabsorption, scattering, and transmission coefficients \citep{Henke93}. We estimated the absorbed energy by multiplying the incident flux by the H$_2$O transmission spectrum, assuming that the incident X-ray photons are primarily absorbed by H$_2$O (a thickness of $\sim$0.24~$\mu$m with a density of 0.87~g~cm$^{-3}$; \citealt{Dohnalek03,Bouilloud15}). This estimate takes into account the total energy absorbed in the entire  H$_2$O ice since X-ray photons can penetrate the entire ice/dust sample and generate photoelectrons and Auger electrons along their penetration paths. However, the studied reactions are expected to be triggered by the electrons generated near the interface regions. The relevant distance of these electrons depends on the electron energy \citep{Valkealahti1989, Dupuy2020}. We used CASINO (Monte Carlo simulation of electron trajectory in solids\footnote{https://www.gegi.usherbrooke.ca/casino/}) to estimate the penetration (transferring) depth and the distribution of deposited energy in Fig. \ref{FigA4}. The transferring range of Auger/photoelectrons spans between $\sim$10 and $\sim$150 monolayers (1 ML=1~$\times$~10$^{15}$ molecules cm$^{2}$) depending on the electron energy. Therefore, it is important to note that this estimation only provides an approximation. Since the photon flux is dependent on the photon energy (or photon wavelength), the total photon flux is derived by integrating the individual photon fluxes (as a function of photon energy) over a photon energy range (250-1250 eV) as shown in Fig.~\ref{FigA1}. The same is also applied to estimate the total absorbed energy flux. The photon (energy) fluence is the product of the total photon (energy) flux and time.

\section{Product evolution}\label{appendix_C}

\begin{figure}[t]
	\begin{center}
		\includegraphics[width=85mm]{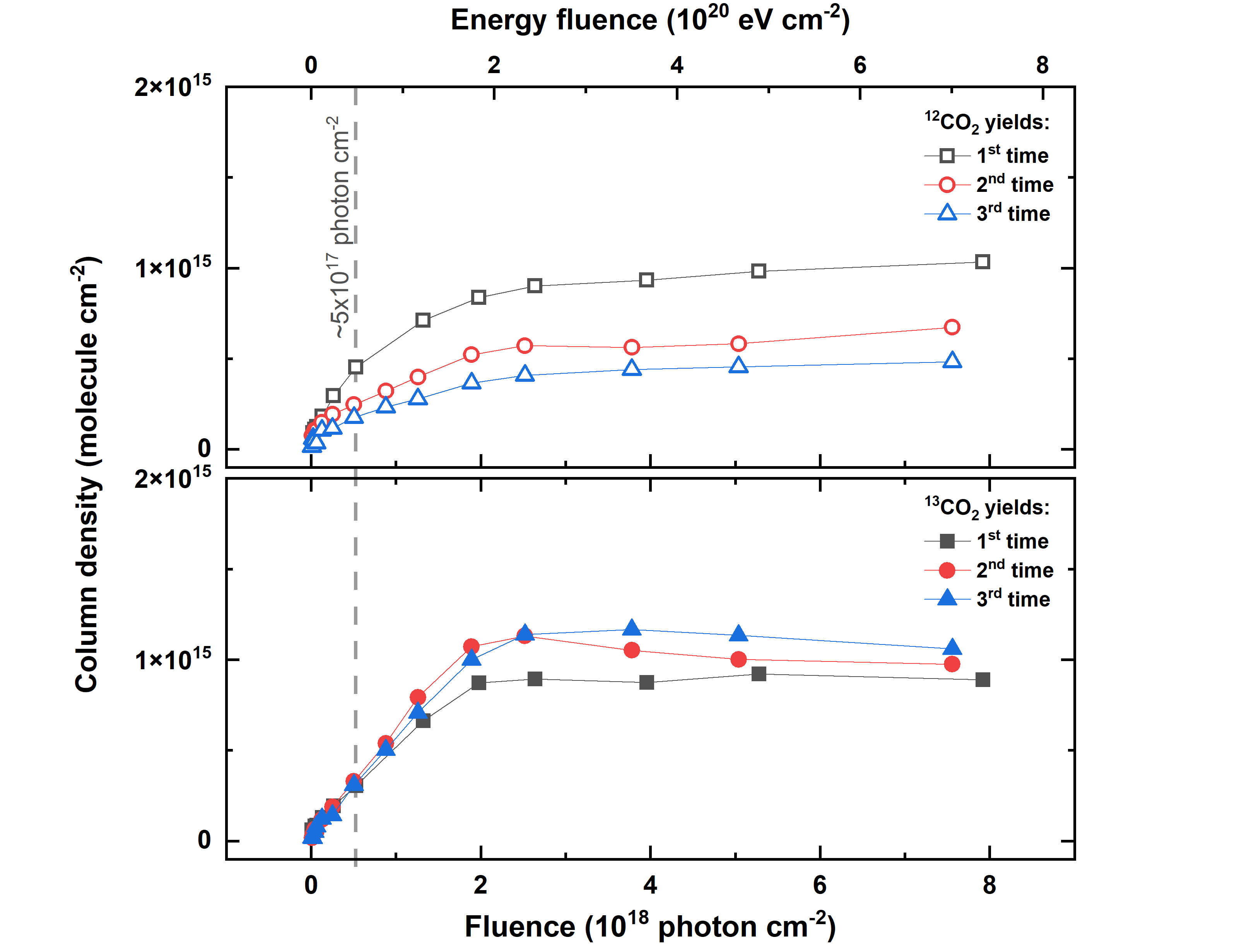}
		\caption{Abundance evolution of the product $^{12}$CO$_2$ (upper) and $^{13}$CO$_2$ (bottom) obtained from the X-ray irradiation of H$_2$O ice on the same a-$^{13}$C dust over an X-ray photons fluence of (7.6--8.0)~$\times$ 10$^{18}$ photon cm$^{-2}$ (i.e., (7.0--7.4)~$\times$ 10$^{20}$~eV~cm$^{-2}$). The dashed line indicates the photon fluence of $\sim$5~$\times$~10$^{17}$ photon cm$^{-2}$. The solid lines are only present as guidance.}
		\label{FigA3}
	\end{center}
\end{figure}

Figure \ref{FigA3} shows the kinetic evolution of $^{12}$CO$_2$ (upper) and $^{13}$CO$_2$ (bottom) as a function of the incident photon fluence or absorbed energy, respectively. The final yield of $^{12}$CO$_2$ decreases over three sequential experiments; the derived abundances are $\sim$10~$\times$~10$^{14}$, $\sim$7~$\times$~10$^{14}$, and $\sim$5~$\times$~10$^{14}$ molecule cm$^{-2}$ for the first, second, and third experiments, respectively. This supports that the $^{12}$C-bearing products result from the surface reactions involving adsorbate contamination enriched in $^{12}$C over $^{13}$C. In contrast, the final $^{13}$CO$_2$ yield  remains relatively constant; $\sim$9~$\times$~10$^{14}$, $\sim$10~$\times$~10$^{14}$, and $\sim$~11$\times$~10$^{14}$ molecule cm$^{-2}$ for the first, second, and third experiments, respectively.

\bibliography{ref}
\bibliographystyle{aasjournal}



\end{document}